\documentclass[sigconf,10pt,nonacm]{acmart}

\usepackage{amsmath,amsfonts}
\usepackage{graphicx}
\usepackage{textcomp}
\usepackage{xcolor}
\usepackage{lipsum}
\usepackage{kotex}
\usepackage{subfig}
\usepackage{algorithm}
\usepackage{algpseudocode}
\usepackage{xurl}
\usepackage{enumitem}
\usepackage{wrapfig}
\usepackage{xspace}
\usepackage{booktabs}
\setcopyright{none}
\renewcommand\footnotetextcopyrightpermission[1]{}

\begin{document}

\title{Revisiting Index Performance\\
under High Cache-Miss Penalties}

\author{Matthew M. Kim}
\affiliation{%
    \institution{Jones College Prep}
    \city{Chicago}
    \state{Illinois}
    \country{USA}
}

\author{Beomseok Nam}
\affiliation{%
    \institution{Sungkyunkwan University}
    \city{Suwon}
    \country{South Korea}
}

\begin{abstract}

Modern data-intensive applications often query very large datasets, making index
performance critical. Recent index structures exploit the fine-grained access
capabilities of emerging storage devices such as DCPMM and CXL. However, existing
benchmarks typically evaluate indexes in isolation using saturated, closed-loop
workloads that emphasize maximum throughput. Such workloads favor highly
concurrent designs while overlooking the application work that follows an index
lookup and competes for cache, memory bandwidth, and CPU resources.
We present {\it AuxBench}---a microbenchmark that evaluates indexes as
\emph{auxiliary access paths} within larger application workloads. Using AuxBench, we evaluate four byte-addressable indexes under
controlled post-lookup memory activity, concurrency, and request
arrival patterns. Our experiments show that post-lookup memory
activity can change the relative performance of competing indexes,
that concurrency-oriented designs may lose their advantages when
substantial application work is introduced, and that time-varying
request arrivals can produce severe tail latency near saturation.
These findings demonstrate why index-only benchmarks may not
adequately predict performance in larger application workloads.

\end{abstract}

\maketitle
\pagestyle{plain}

\sloppy

\section{Introduction}

The emergence of byte-addressable memory technologies, such as Intel Optane
DCPMM~\cite{xiang2022characterizing} and Compute Express Link (CXL)
memory~\cite{cxl}, has motivated numerous index designs that exploit
fine-grained memory access~\cite{hwang18fast,liu20lb+trees,Li10fdtree,
chen15,oukid16,arulraj2018bztree}. However, the relatively high access latency
of these technologies makes index performance particularly sensitive to CPU
cache behavior ~\cite{Liu25asplos,Liu25osdi,bando25}. Consequently, many index
designs focus on improving cache locality and reducing cache misses. Their
effectiveness, however, depends not only on the locality of index accesses but
also on the cache capacity available to the index.

Existing studies typically evaluate byte-addressable indexes using isolated,
saturated workloads that maximize index throughput
~\cite{lersch2019evaluating,he22}. Such evaluations implicitly assume that most
CPU and cache resources are available to the index. In real applications,
however, an index is only one component of a larger execution pipeline. For
example, a key--value store accesses the retrieved object after an index lookup,
while a filesystem performs additional metadata operations. These activities
compete with index operations for cache capacity, memory bandwidth, and CPU
resources. As a result, cache-conscious index designs that perform well in
isolation may lose their advantages when application accesses evict frequently
used index nodes from CPU caches. Furthermore, saturated workloads continuously
exercise the index, favoring highly concurrent designs whose synchronization
optimizations may provide smaller benefits when application work is interleaved
with index operations.

These limitations motivate evaluating indexes as auxiliary components of
application workloads rather than as isolated data structures. We introduce
{\it AuxBench}, a microbenchmark that integrates existing index implementations
with configurable post-lookup memory activity and independently controlled
request arrivals. This allows us to examine how application work, concurrency,
and request dynamics affect the relative performance of competing indexes.

The contributions of this work are as follows:
\begin{itemize}[leftmargin=3.0mm]

\item \textbf{Application-aware index benchmarking.}
We introduce {\it AuxBench}, a microbenchmark that evaluates indexes 
as auxiliary components of application workloads. Its common
interface allows indexes to be evaluated under
the same configurable post-lookup memory activity, request arrival
patterns, and operation mixes.

\item \textbf{Cache interference and index performance.} We demonstrate that
application-induced cache interference affects index designs differently,
changing their relative performance and, in some cases, reversing their
rankings. In particular, cache-conscious designs do not necessarily retain their
advantages under heavy cache pollution.

\item \textbf{Workload-dependent concurrency benefits.} We show that the
scalability advantages of highly concurrent index designs can diminish
substantially when application work is interleaved with index operations. We
also observe significant differences in how index implementations respond to
write-intensive workloads.

\item \textbf{Request dynamics and tail latency.} Using open-loop and
time-varying workloads, we demonstrate how differences in index processing
capacity translate into substantial differences in queueing delay and
high-percentile response latency under externally controlled demand.

\end{itemize}

We evaluate AuxBench using four existing byte-addressable index
structures---FAST\&FAIR~\cite{hwang18fast}, WB+Tree~\cite{chen15}, 
$\mu$Tree~\cite{chen20}, and FPTree~\cite{oukid16}. Our results show that
index performance rankings and the benefits of concurrency-oriented
optimizations can change substantially when the surrounding application workload
is considered. These findings highlight the importance of evaluating indexes not
only by their peak throughput in isolation but also by their performance under
application-induced cache interference and realistic request patterns.

\section{Benchmark Design}
\label{sec:design}

\begin{figure}
    \centering
    \includegraphics[width=1.0\linewidth]{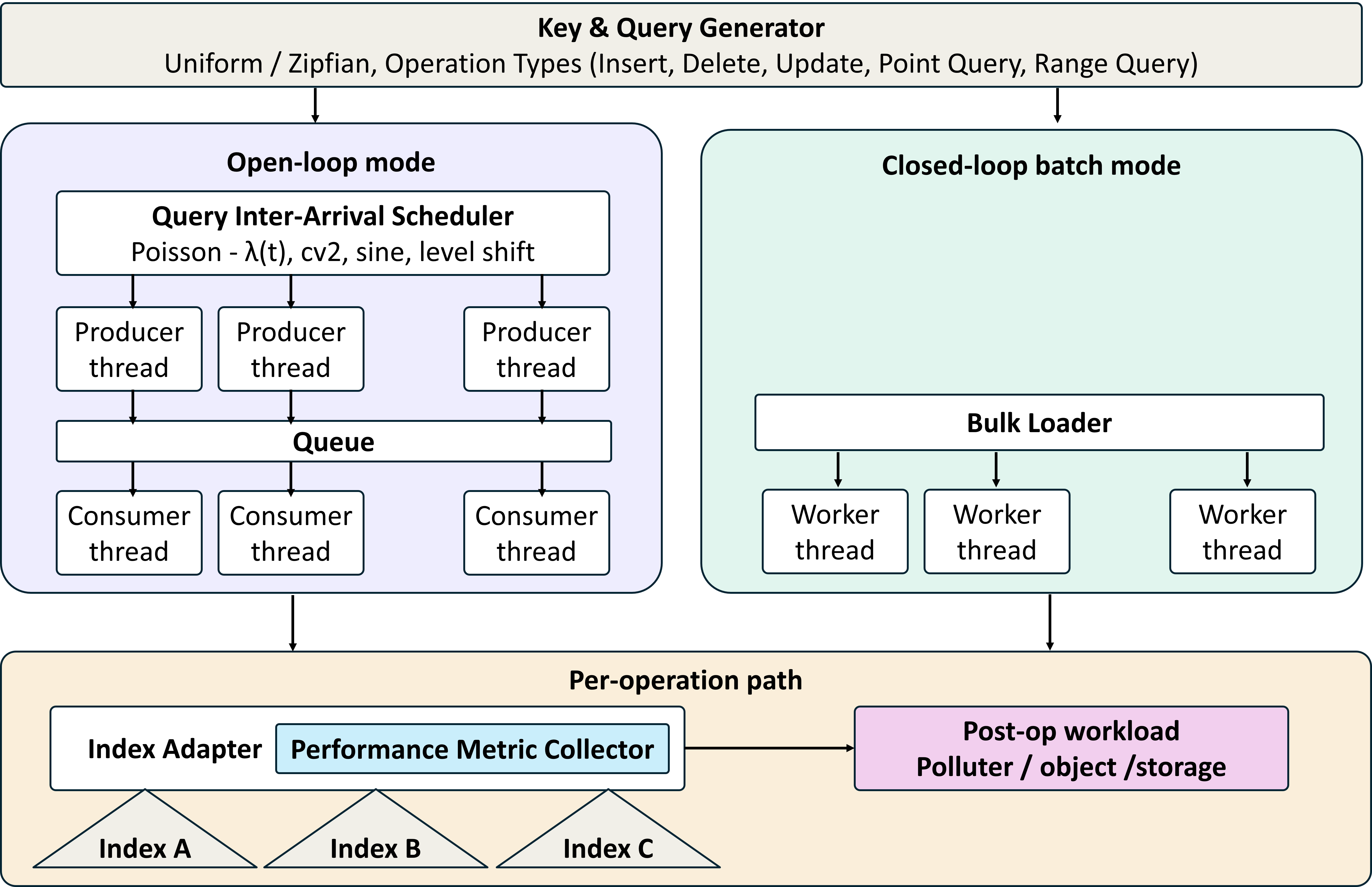}
    \caption{Architecture of AuxBench}
    \label{fig:arch}
\end{figure}

AuxBench is a microbenchmark for evaluating byte-addressable indexes
under configurable application-level workloads. As illustrated in
Figure~\ref{fig:arch}, existing index implementations are integrated
through a common interface, while the benchmark independently controls
the surrounding workload. This separation allows different indexes to
be evaluated under identical workload configurations without modifying
their internal data structures or concurrency mechanisms.

AuxBench supports three main workload dimensions: (i) post-lookup
memory activity, which introduces configurable memory accesses after
index operations; (ii) request arrival patterns, which control the
rate and temporal distribution of requests; and (iii) operation mixes,
which determine the proportions of searches, updates, inserts, deletes,
and scans. These dimensions can be configured independently of the
index implementation and one another.

The objective is to evaluate how application-level workloads affect
the relative performance of competing indexes, rather than simply
measuring their isolated performance. For example, an index optimized
for cache locality may lose its advantage when application memory
accesses compete for cache capacity. Similarly, the benefits of
concurrency-oriented optimizations may diminish when additional
application work reduces contention between index operations.

\subsection{Post-Lookup Workload}
\label{sec:postlookup}

The distinguishing feature of AuxBench is its ability to control the memory
activity performed between consecutive index operations. In conventional
microbenchmarks, a worker typically performs index operations back-to-back.
Consequently, frequently accessed index nodes may remain resident in CPU caches,
and the measured performance can depend heavily on the cache locality
established by the benchmark itself. In an application, however, an index lookup
is usually followed by additional work, such as accessing a retrieved object,
processing a key-value pair, or traversing metadata. These activities consume
cache capacity and memory bandwidth, potentially changing the cache behavior of
subsequent index operations. AuxBench models such interference by executing a
configurable workload after each index operation.  The workload is independent
of the index implementation, allowing application memory activity to be varied
without changing the index or its input key distribution.

AuxBench supports the following post-lookup modes:
\begin{itemize}[leftmargin=3.0mm]
    \item \textbf{None:}
    No additional work is performed after an index operation.
    This corresponds to conventional index-only benchmarking
    and serves as the baseline.

    \item \textbf{ObjectAccess:}
    The benchmark reads a contiguous memory region and computes
    a checksum over its contents. This approximates application
    processing after an index lookup, such as reading and
    processing an object in a key-value store. By varying the
    amount of data accessed, AuxBench examines how application-level
    memory activity affects index performance.

    \item \textbf{Polluter:}
    The benchmark accesses and modifies a configurable amount
    of memory from a separate buffer in a randomly permuted
    order. Unlike ObjectAccess, which models sequential object
    processing, Polluter deliberately introduces less predictable
    memory accesses and write traffic to stress the CPU cache.
    This allows AuxBench to examine how different indexes respond
    to aggressive cache interference.
\end{itemize}

The post-lookup workload is parameterized by the amount of memory accessed per
operation, $N$, and, for Polluter, the size of the underlying buffer, $U$.
These parameters control different aspects of memory interference. Increasing
$N$ introduces more memory activity between index operations, while changing
$U$ alters the working set over which those accesses are distributed.
By sweeping these parameters, AuxBench can evaluate indexes across a range of
cache environments, from index-only execution to workloads in which application
data competes substantially with index nodes for cache capacity.
The distinction between Polluter and ObjectAccess is intentional. Polluter
provides a controlled source of cache interference, whereas ObjectAccess
approximates the sequential memory accesses that commonly follow an index
lookup. Comparing these modes helps distinguish the effects of application
access patterns on different index designs.
Importantly, AuxBench evaluates the combined execution of an index operation
and its associated post-lookup workload. Therefore, changes in total operation
latency may reflect both the cost of the additional work and its effect on
subsequent index accesses. 

%The evaluation examines how these effects influence
%the relative performance of competing indexes.

\subsection{Request Arrival Patterns}
\label{sec:arrival}

%The rate at which an application invokes an index is another important factor
%in index performance.  

Conventional index microbenchmarks~\cite{he22,lersch2019evaluating} commonly use
closed-loop execution, in which each worker issues a new operation immediately
after completing the previous one. In this approach, the request rate is
determined by the execution speed of the index itself. In real applications,
however, request arrivals must be independent of index lookup time.

AuxBench supports both closed-loop and open-loop execution. In open-loop mode,
requests are generated according to an externally specified arrival process,
independently of the completion of previous operations.  A dedicated producer
places requests into a queue, from which consumer threads retrieve and execute
them.

This separation enables different indexes to be evaluated under the same
offered request rate, rather than allowing each index to determine its own
arrival rate through its execution speed.

\paragraph{Inter-arrival variability.}

AuxBench controls not only the average request rate but also the variability of
inter-arrival times.  For a target arrival rate $\lambda$, inter-arrival times
are sampled from a Gamma distribution:
\begin{equation}
T_{\mathrm{ia}} \sim
\operatorname{Gamma}
\left(
k=\frac{1}{CV^2},
\theta=\frac{1}{\lambda k}
\right),
\label{eq:interarrival}
\end{equation}
where $CV$ is the coefficient of variation of the inter-arrival time
distribution.

When $CV^2=1$, the distribution is exponential, corresponding to Poisson
arrivals at a constant rate.  Smaller values produce more regular arrivals,
while larger values produce increasingly variable arrivals.
This parameterization allows workloads with the same average request rate but
different degrees of burstiness to be compared without changing the index or
key distribution.

\paragraph{Time-varying arrivals.}

AuxBench also supports workloads whose request rates change over time. The
target rate is expressed as
\begin{equation}
\lambda(t) =
R_0 \cdot \operatorname{level}(t)
\cdot
\left(
1+A\sin\left(\frac{2\pi t}{P}+\phi\right)
\right)
\cdot \operatorname{burst}(t),
\label{eq:arrival_rate}
\end{equation}
where $R_0$ is the baseline arrival rate.  The level component represents
piecewise-constant changes, the sinusoidal component represents periodic
variation, and the burst component introduces short-lived increases in request
intensity.

These patterns are not new workload models.  AuxBench incorporates them to
examine how different index implementations behave when the offered load varies
over time.

\paragraph{Latency measurement.}

Open-loop execution distinguishes operation execution time from queueing delay.
For each request, AuxBench records its enqueue time, execution start time, and
completion time:
\begin{align}
T_{\mathrm{queue}}
    &= t_{\mathrm{start}} - t_{\mathrm{enqueue}},\\
T_{\mathrm{service}}
    &= t_{\mathrm{end}} - t_{\mathrm{start}}.
\end{align}
Here, service time includes both the index operation and the 
post-lookup workload. Queueing delay captures the time a request waits before
execution.  AuxBench reports these distributions separately, including
high-percentile latencies.

The benchmark also records producer lag when the request generator cannot
maintain the specified arrival schedule. This allows experiments in which the
offered load is not successfully delivered to be identified.
These measurements are useful when comparing indexes near saturation, where
relatively small differences in processing capacity can result in
substantial differences in end-to-end response latency.

\subsection{Operation Mix}
\label{sec:operation_mix}

%Index performance also depends on the types of operations executed
%concurrently. Read-only workloads primarily exercise lookup paths, whereas
%updates, inserts, and deletes can introduce synchronization, cache-line
%modifications, and additional memory traffic.  Range scans may further affect
%cache behavior by accessing multiple consecutive entries.

AuxBench supports five operation types: search, update, insert, delete, and
scan. Their relative frequencies are controlled through a normalized
probability vector:
\begin{equation}
(s,u,i,d,sc), \qquad s+u+i+d+sc=1.
\end{equation}
The operation mix can be changed independently of the request arrival process
and post-lookup workload.  

%Consequently, the same index can be compared under read-dominated, write-heavy,
%and scan-heavy workloads while keeping the remaining experimental conditions
%fixed.

AuxBench provides built-in reference indexes as well as an interface for
integrating existing third-party indexes. External implementations retain their
own concurrency-control and index-maintenance mechanisms.  This is important
because the goal is to compare the behavior of index designs, not to
impose a common synchronization strategy on them.
The built-in indexes use simplified maintenance operations. In particular, the
reference B+-tree models insertion as an upsert and deletion as a tombstone
write, rather than implementing full node splitting and structural maintenance.
These implementations provide reference baselines and are not intended to
characterize the complete maintenance costs of production B+-trees.
For every operation type, AuxBench collects latency distributions separately,
including mean, median, and high-percentile latencies. This prevents the
performance of less frequent but potentially expensive operations from being
obscured by aggregate throughput or overall average latency.
Together, the configurable operation mix and post-lookup workload allow
AuxBench to examine how the relative benefits of different index designs change
when concurrency, cache interference, and memory-modification activity are
introduced.

\section{Implementation}\label{sec:implementation}

AuxBench integrates index implementations through a common interface. The
benchmark manages request generation, workload execution, and performance
measurement independently of the index being evaluated. This separation allows
the same workload and measurement infrastructure to be used across different
index designs. The implementation focuses on minimizing interference introduced
by the benchmark itself, particularly in open-loop execution, where request
generation and processing proceed concurrently.

\subsection{Request Generation and Execution}
\label{sec:request_execution}

In closed-loop mode, each worker generates and executes index operations
sequentially, issuing its next request after the previous one completes. In
open-loop mode, a dedicated producer generates requests according to the
configured arrival process and places them into a bounded queue. Consumer
threads independently dequeue requests and execute the corresponding index
operations and post-lookup workloads.

The producer maintains an absolute arrival schedule. For each request, it
computes the next target arrival time and waits until that time before
attempting to enqueue the request. Unlike a relative-delay approach, this
prevents request-generation overhead from accumulating into systematic drift in
the arrival schedule.

%The queue capacity is a power of two, and producer and consumer position
%counters are placed on separate cache lines to %reduce false sharing.

The request queue uses a bounded lock-free multiple-producer multiple-consumer
(MPMC) ring buffer based on Vyukov's algorithm~\cite{vyukov2011bounded}. Each
slot maintains a sequence number to coordinate concurrent enqueue and dequeue
operations using atomic instructions. Although the queue supports multiple
producers, AuxBench uses a single dedicated producer for open-loop request
generation. The MPMC implementation provides a reusable concurrent queue without
requiring the index implementations to participate in request scheduling.

The producer and consumers use busy-waiting rather than blocking
synchronization. To reduce interference from thread migration and scheduling,
each thread is pinned to a dedicated physical CPU core. Consequently, open-loop
execution requires one core for the producer in addition to the cores assigned
to index workers.

For time-varying workloads, the producer evaluates the configured arrival-rate
function and samples inter-arrival times using the Gamma distribution described
in Section~\ref{sec:arrival}. The rate is treated as approximately constant over
each sampled interval. This approximation is intended for workload variations
occurring on millisecond-to-second time scales, rather than for modeling rate
changes at individual instruction timescales.

\subsection{Latency Collection}
\label{sec:latency_collection}

AuxBench records three timestamps for each request: the time at which it is
enqueued, the time at which a consumer begins processing it, and
the time at which processing completes. As defined in Section~\ref{sec:arrival},
these timestamps distinguish queueing delay from service time.
Service time includes both the index operation and the configured post-lookup
workload. Queueing delay captures the time spent waiting for an available
consumer. In closed-loop mode, queueing delay is zero by construction.

To avoid contention in the measurement path, each consumer maintains
thread-local latency histograms. AuxBench uses logarithmically organized
histograms with 40 power-of-two ranges, each divided into eight approximately
linear sub-buckets. 
The histogram resolution is approximately 10\%. In particular, AuxBench
reports the lower boundary of the bucket containing the requested percentile.
The benchmark reports mean, median, p99, p99.9, p99.99, and maximum latency for
each operation type. Separate histograms for queueing delay and service time
allow their contributions to response latency to be examined independently.

\subsection{Measurement Fidelity and Limitations}
\label{sec:measurement_fidelity}

Open-loop benchmarking introduces additional execution overhead because request
generation, queue management, and index processing occur concurrently. AuxBench
minimizes this interference through a lock-free request queue, dedicated CPU
cores, and thread-local latency collection.

Nevertheless, these mechanisms do not eliminate measurement overhead entirely.
Each request incurs queue operations, timestamp collection, and histogram
updates. Furthermore, busy-waiting consumes CPU resources even when the request
rate is low. These costs must be considered when interpreting results,
particularly for indexes whose operations complete in very short periods.

Another limitation arises when the offered request rate exceeds the rate that
the benchmark can deliver. Because the queue is bounded, the producer may be
unable to enqueue a request at its scheduled arrival time. AuxBench records this
delay as producer lag, defined as the difference between the scheduled arrival
time and the actual enqueue time.

Substantial producer lag indicates that the configured arrival process was not
faithfully delivered. Such experiments must be distinguished from cases where
requests are successfully generated at the target rate but experience long
queueing delays because the index cannot process them quickly enough.

Finally, AuxBench measures service time as the combined execution time of an
index operation and its post-lookup workload. It does not separately attribute
this time to the index and application work. Consequently, measurements of
service-time degradation alone cannot determine whether the change originates
from additional application execution or from increased index cache misses.

The evaluation therefore emphasizes comparisons between index implementations
under identical post-lookup workloads, rather than interpreting the combined
service time as the intrinsic latency of an individual index.

\section{Experimental Evaluation}
\label{sec:evaluation}

We use AuxBench to investigate whether the relative performance of
byte-addressable indexes observed in conventional microbenchmarks persists when
the indexes are evaluated as components of larger application workloads. Rather
than evaluating each workload parameter in isolation, we focus on three
questions:

\begin{itemize}[leftmargin=3.0mm]
    \item \textbf{Cache interference:}
    Does application memory activity change the relative
    performance of indexes, and how does this effect interact
    with access patterns and key locality?

    \item \textbf{Concurrency and updates:}
    Do the benefits of concurrency-oriented index designs
    persist when application work is interleaved with index
    operations, or when writes introduce additional contention?

    \item \textbf{Request dynamics:}
    How do differences in index processing capacity affect
    response latency under externally controlled and
    time-varying request arrivals?
\end{itemize}

These experiments are intended to identify conditions under which conclusions
drawn from index-only benchmarks may not generalize to application workloads.

\subsection{Experimental Setup}
\label{sec:eval_setup}

\paragraph{Testbed.} Our experiments are conducted on a dual-socket server with
two Intel Xeon Gold 5215 processors running at 2.50 GHz, 256 GB of DRAM, and 1
TB of Intel Optane DCPMM (4 DCPMM per socket) in app-direct mode.  The server
runs Ubuntu 22.04 with Linux kernel 6.8.0-40-generic.

\paragraph{Indexes.} We evaluate four representative byte-addressable B+-tree
indexes: WB+Tree, $\mu$Tree, FAST\&FAIR, and FPTree.  These indexes were
designed with different approaches to cache utilization, persistence, and
concurrency control. Each index is integrated into AuxBench~\footnote{The
source code of AuxBench will be released upon publication.} through the common
index interface, allowing the same workload configuration to be applied without
changing the underlying index implementation.

\paragraph{Workloads and metrics.} Unless otherwise specified, each experiment
varies one workload parameter while keeping the remaining parameters fixed. We
consider the amount and access pattern of post-lookup memory activity, key
distribution, worker concurrency, operation mix, and request arrival patterns.
For closed-loop experiments, we primarily report throughput or operation
latency. For open-loop experiments, we distinguish service time from queueing
delay and examine high-percentile end-to-end latency. Our primary interest is
not only the absolute performance of an individual index but also whether its
relative advantage over competing designs changes across workload conditions.

%\paragraph{Experimental scope.}
%Our experiments use Intel Optane DCPMM rather than CXL-attached
%memory. The results characterize the evaluated indexes on this
%testbed and should not be interpreted as direct measurements of
%CXL performance. CXL motivates the broader importance of
%examining indexes under competing memory activity; evaluation
%on CXL hardware remains future work.

\begin{figure}
    \centering
    \begin{minipage}{0.48\linewidth}
        \includegraphics[width=1.0\linewidth]{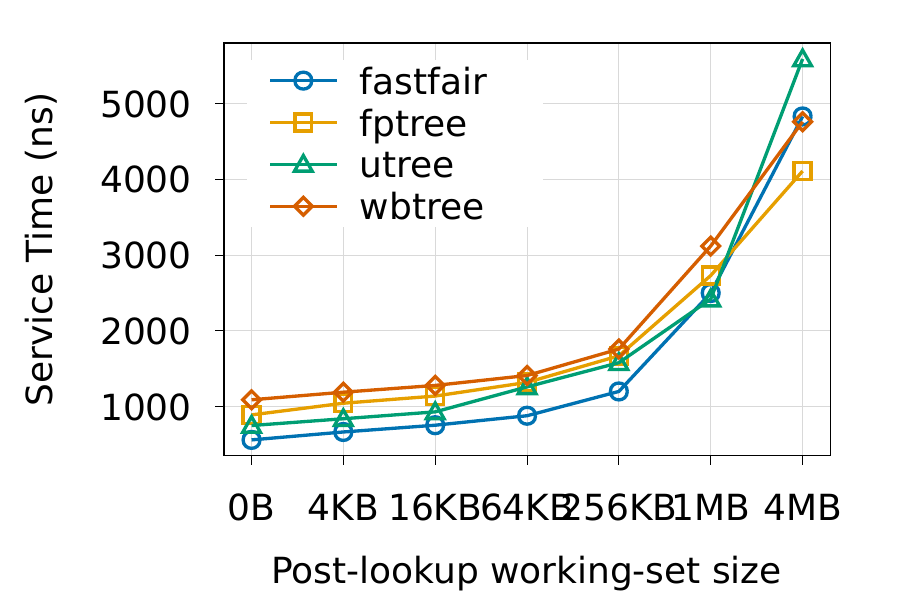}
        \caption{Post-lookup working set size affects the relative performance of indexes}
        \label{fig1}
    \end{minipage}
    \hfill
    \begin{minipage}{0.48\linewidth}
        \includegraphics[width=1.0\linewidth]{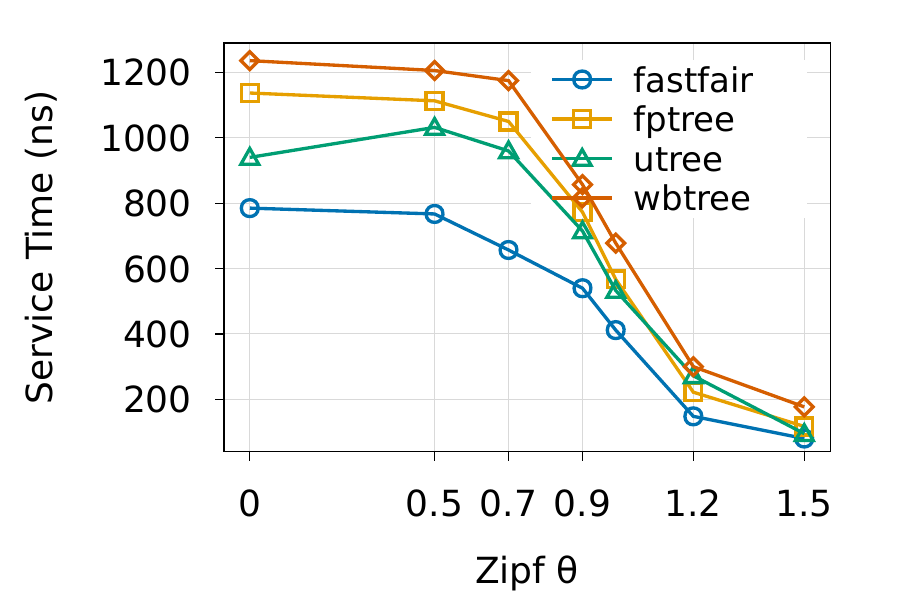}
        \caption{Key skew changes cache locality and relative performance}
        \label{fig3}
    \end{minipage}
\end{figure}

\subsection{Cache Interference and Performance}
\label{sec:eval_cache}

We first examine how memory accesses performed by the surrounding application
affect the performance of indexes. In an index-only benchmark,
repeated index operations can benefit from cache residency established by
previous lookups. When an application accesses other data between lookups, the
cache environment changes, potentially affecting indexes differently.
We study this interaction by varying the amount of post-lookup memory activity,
its access pattern, and the locality of index requests.

\subsubsection{Impact of Post-Lookup Working-Set Size}

Figure~\ref{fig1} compares the indexes as the amount of
post-lookup memory activity increases.  With no post-lookup work, the
experiment corresponds to a conventional index-only microbenchmark. Increasing
the pollution size introduces additional memory accesses between consecutive
index operations.
As expected, larger post-lookup workloads generally increase execution time.
More importantly, however, the degradation is not uniform across indexes. The
performance gaps between the evaluated indexes change substantially as the
amount of application memory activity increases.

A particularly interesting observation concerns $\mu$Tree.  Although $\mu$Tree
is designed to improve cache locality, it exhibits the worst performance among
the evaluated indexes under severe cache pollution. This suggests that
cache-oriented optimizations that are effective under index-only workloads do
not necessarily provide the same benefit when index data competes with
application data for cache capacity. The observed performance degradation of $\mu$Tree may be partly
explained by its hybrid organization, which combines a
DRAM-resident B$^+$-tree with a persistent-memory linked list. Although this design improves cache locality
under index-only workloads, cache pollution can evict frequently accessed
in-memory nodes and expose the cost of dependent memory accesses. 

The experiment reveals more than the expected cost of additional memory
accesses. It shows that indexes have different sensitivities to the cache
environment, and that these differences can change their relative performance.
This result has an important implication for index evaluation. A benchmark that
repeatedly accesses an index without intervening application work may favor
designs whose performance depends strongly on cache residency.  Such an
evaluation does not necessarily identify the index that performs best when
deployed within an application.

\subsubsection{Interaction with Key Locality}

The cache environment experienced by an index depends not only on application
memory activity but also on the locality of index requests. We therefore vary
the key distribution from uniform to increasingly skewed accesses while
retaining the post-lookup workload.

Figure~\ref{fig3} shows that increasing key skew reduces lookup latency
by concentrating accesses on a smaller subset of index entries. This improves
the likelihood that frequently accessed index nodes remain resident in cache,
even when application work introduces additional memory activity.

The resulting reduction in absolute latency is not uniform across indexes.
Although the absolute latency gaps become smaller under higher skew, the
relative speedups can increase. Thus, improvements in cache locality do not
necessarily preserve the relative performance differences observed under
uniform access.

These observations demonstrate that application memory activity and key
locality interact in determining index performance. Evaluating either dimension
alone may therefore produce an incomplete characterization of the cache
behavior of an index.

\begin{figure}
    \centering
    \includegraphics[width=1.0\linewidth]{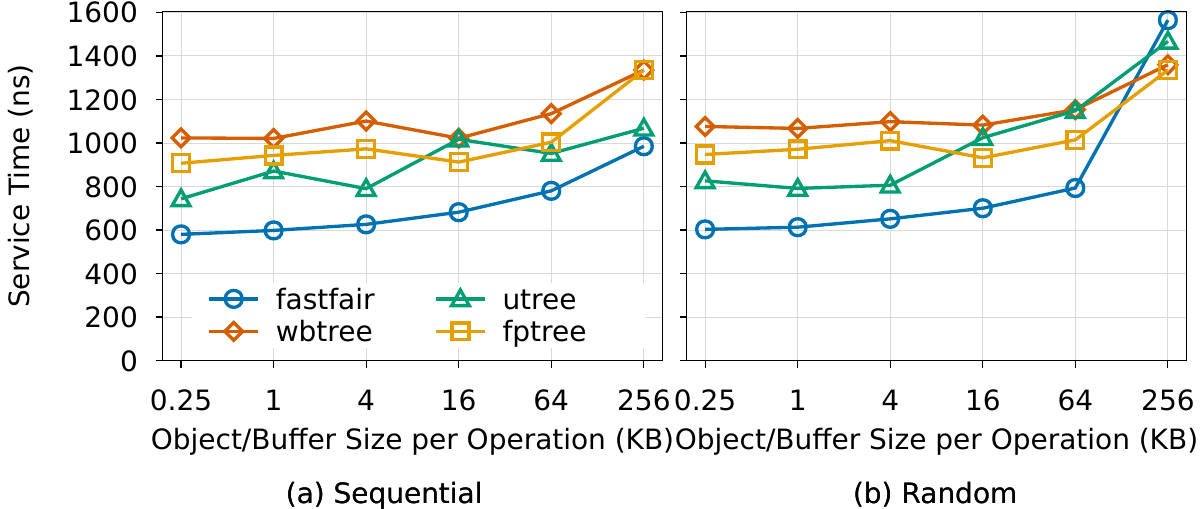}
    \caption{Post-lookup access patterns affect service time differently across indexes}
    \label{fig4}
\end{figure}

\subsubsection{Impact of Application Access Patterns}

We next examine whether different application access patterns produce similar
effects. Figure~\ref{fig4} compares sequential and random post-lookup memory
accesses while varying the amount of data accessed per operation. Both access patterns increase operation latency as the amount of data accessed per operation grows. However, the degree of degradation differs among the evaluated indexes.
In particular, FAST\&FAIR exhibits greater performance degradation
under random post-lookup accesses. Its pointer-dependent lookup paths
may contribute to this behavior, although the experiments do not
isolate their effect.

These results indicate that the amount of application memory activity alone is
insufficient to characterize its interference with an index. The access pattern
also matters, and its effect depends on the index implementation.

Taken together, these experiments show that the relative performance of
byte-addressable indexes depends on the cache environment established by the
surrounding application. Cache pollution, key locality, and application access
patterns cannot always be treated as independent factors when comparing index
designs.

\subsection{Application Work and Concurrency}
\label{sec:eval_concurrency}

Byte-addressable indexes often employ fine-grained or lock-free synchronization
to improve throughput under concurrent access. Their benefits are commonly
measured using saturated workloads in which worker threads repeatedly execute
index operations with little or no intervening application work.
Such workloads maximize the frequency of concurrent accesses to the index. In
an application, however, workers may perform substantial computation or memory
accesses between index operations. This changes both the fraction of execution
time spent in the index and the frequency of contention among workers.

\subsubsection{Scalability under Post-Lookup Work}

\begin{figure}
    \centering
    \includegraphics[width=1.0\linewidth]{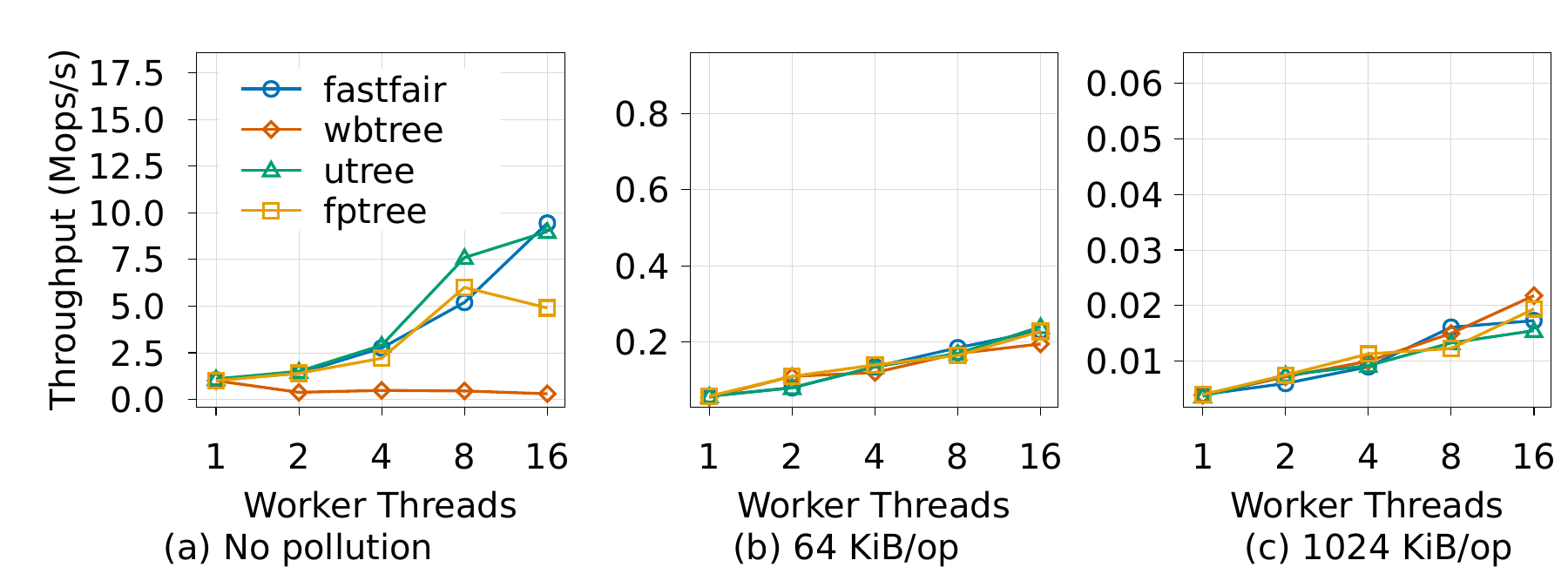}
    \caption{Post-lookup memory activity reduces the observed
scalability advantages of concurrent indexes}
    \label{fig2}
\end{figure}

Figure~\ref{fig2} compares index throughput as worker concurrency increases
under different amounts of post-lookup work. Without post-lookup work,
concurrent workers frequently access the index, allowing synchronization
overhead and differences in concurrency-control mechanisms to play an important
role in scalability. As a result, the throughput of all indexes employing
fine-grained synchronization increases with the number of worker threads. Since
wB+Tree was originally designed for single-threaded execution, a coarse-grained
lock is used to protect the entire index. Consequently, its throughput does not
improve as the number of worker threads increases.

The advantage of highly concurrent index designs becomes substantially smaller
when post-lookup work is introduced. In particular, with 64 KiB and 1 MiB of
cache-polluting accesses per operation, the scalability gains observed in
index-only workloads are greatly reduced.

Two effects can contribute to this behavior.  First, the additional memory
activity consumes cache capacity and memory bandwidth, reducing the fraction of
total execution time attributable to index operations. Second, application work
increases the interval between consecutive index accesses, which can reduce
contention among concurrent workers. As a result, the benefits of fine-grained
or lock-free synchronization may become less visible when the index is embedded
in an application workload.
The experiment illustrates an important limitation of evaluating
concurrency-oriented indexes using only saturated index operations. A design
that scales particularly well under continuous index access does not
necessarily provide a comparable end-to-end benefit when application work
dominates the execution pipeline.

\subsubsection{Impact of Write Operations}

We next examine how the operation mix changes the relative performance of
indexes. Unlike read-only workloads, updates modify shared memory locations and
may introduce additional synchronization and cache-coherence traffic.

Figure~\ref{fig5} compares FAST\&FAIR and $\mu$Tree as the fraction of write
operations increases. These two indexes are more sensitive to cache pollution
than the other indexes, as shown in Figure~\ref{fig4}. They exhibit different
sensitivities to writes. FAST\&FAIR experiences up to a 54$\times$ throughput
degradation, whereas $\mu$Tree experiences approximately a 5$\times$
degradation.

\begin{figure*}[t]
    \begin{minipage}{0.32\textwidth}
        \centering
        \includegraphics[width=1.0\linewidth]{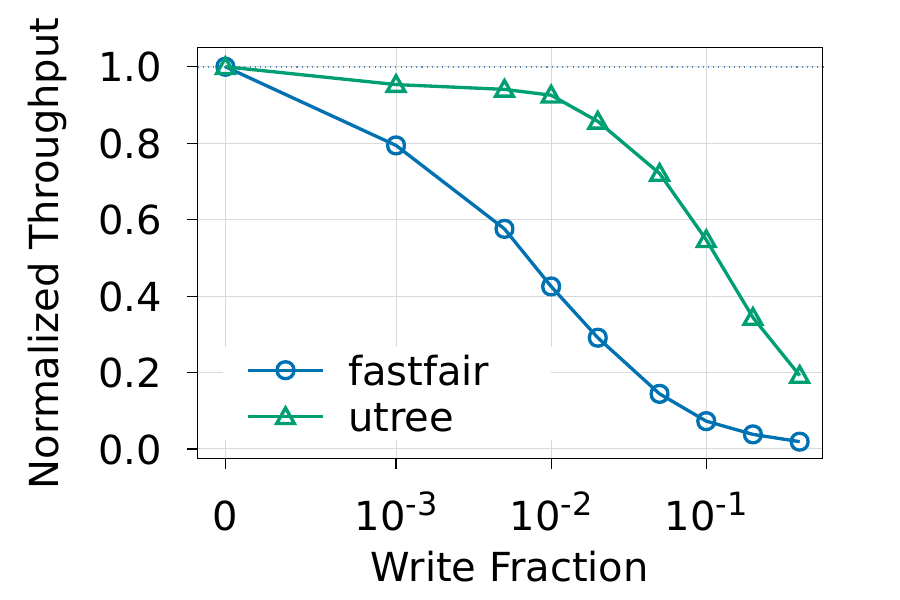}
        \caption{Write fraction reveals index-specific throughput degradation}
        \label{fig5}
    \end{minipage}
    \hfill
    \begin{minipage}{0.32\textwidth}
        \centering
        \includegraphics[width=1.0\linewidth]{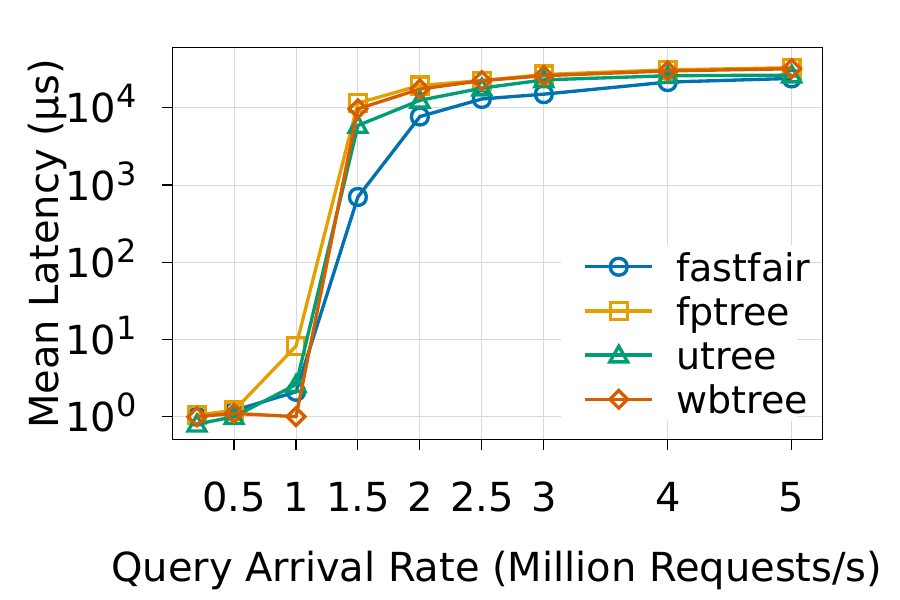}
        \caption{Near saturation, queueing—not service time—dominates response latency}
        \label{fig6}
    \end{minipage}
    \hfill
    \begin{minipage}{0.32\textwidth}
        \centering
        \includegraphics[width=1.0\linewidth]{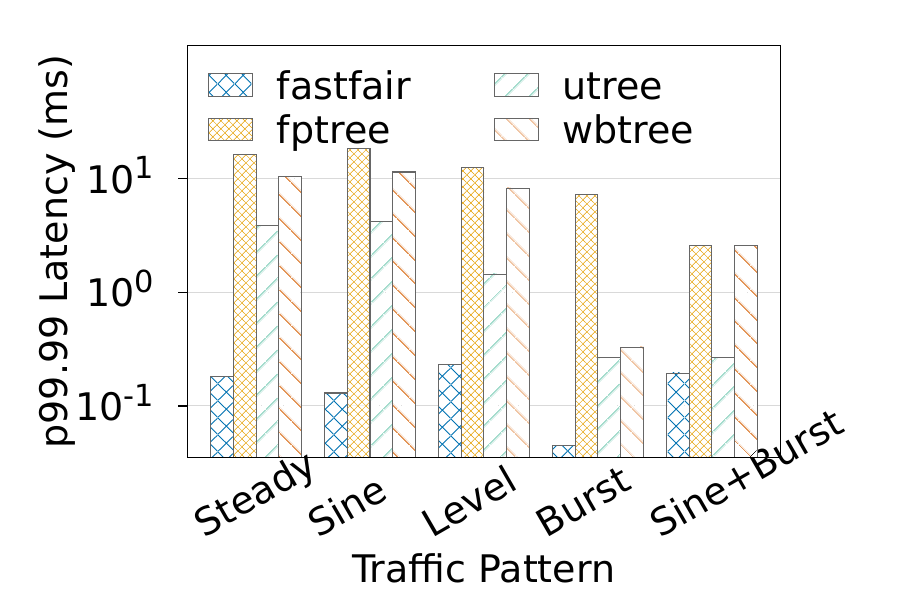}
        \caption{Demand shape changes p99.99 end-to-end latency}
        \label{fig7}
    \end{minipage}
\end{figure*}

\if 0
\begin{figure}[t]
    \centering
    \includegraphics[width=1.0\linewidth]{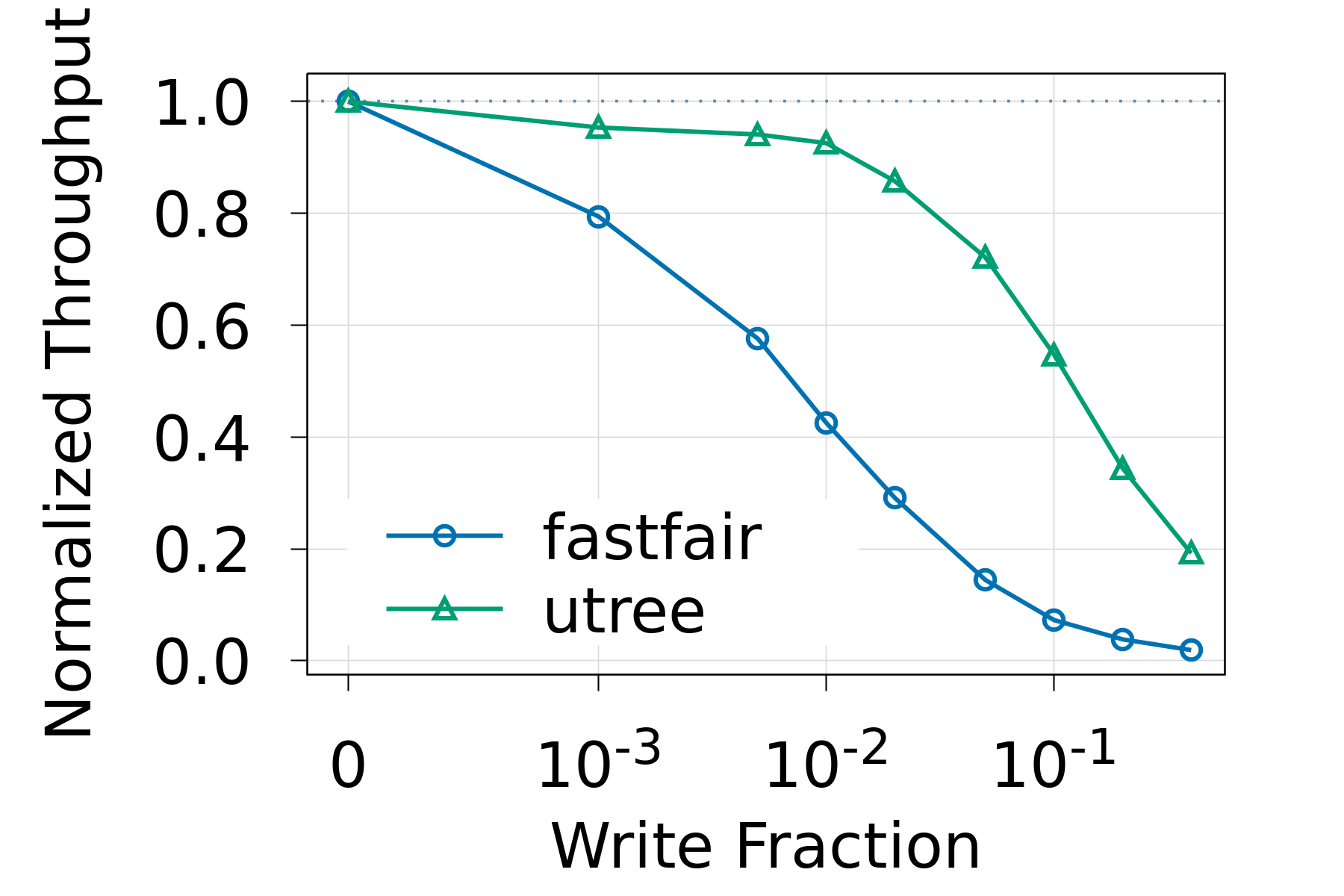}
    \caption{Write fraction reveals index-specific throughput degradation}
    \label{fig5}
\end{figure}

\begin{figure}
    \begin{minipage}{0.48\linewidth}
        \centering
        \includegraphics[width=1.0\linewidth]{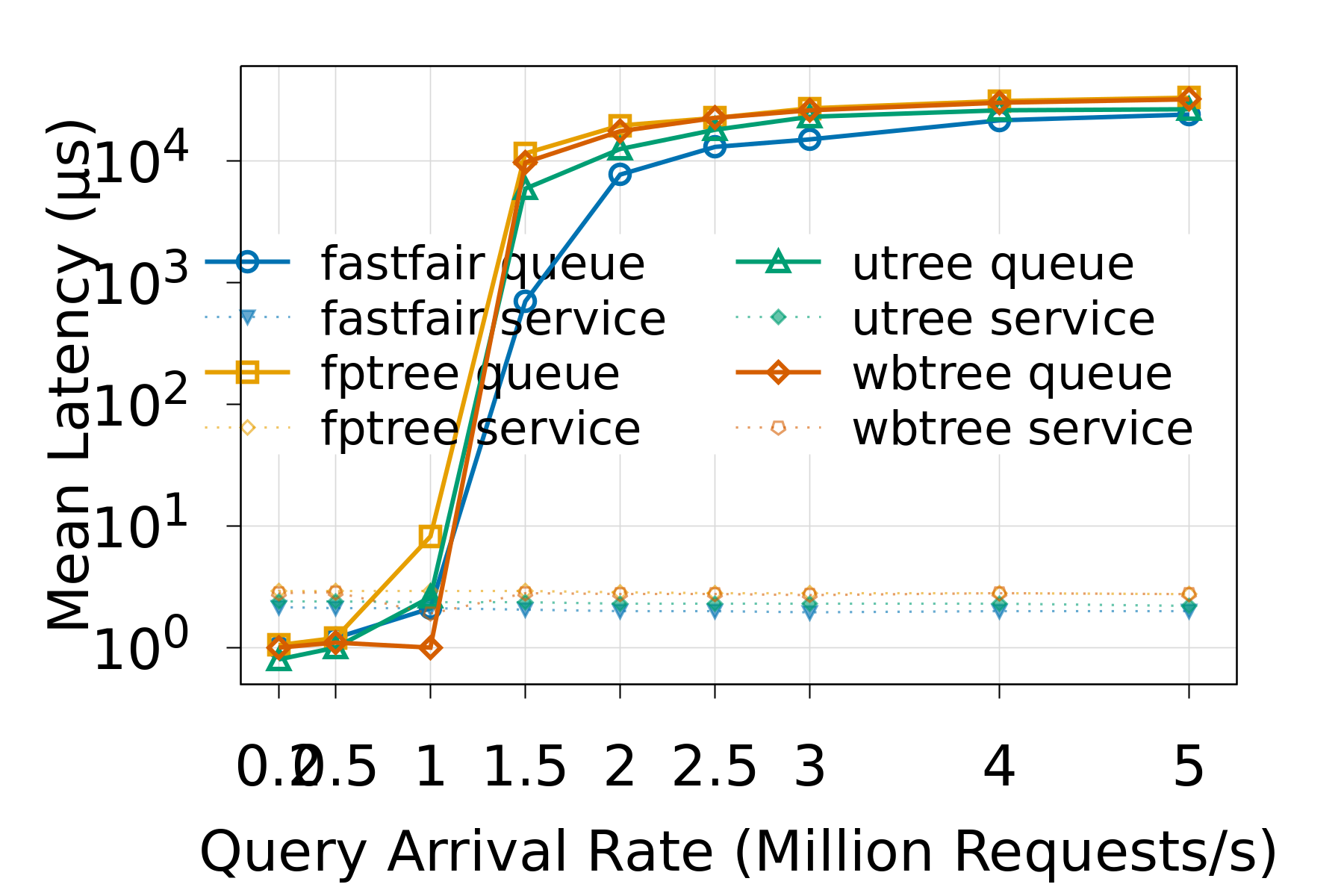}
        \caption{Near saturation, queueing—not service time—dominates response latency}
        \label{fig6}
    \end{minipage}
    \begin{minipage}{0.48\linewidth}
        \centering
        \includegraphics[width=1.0\linewidth]{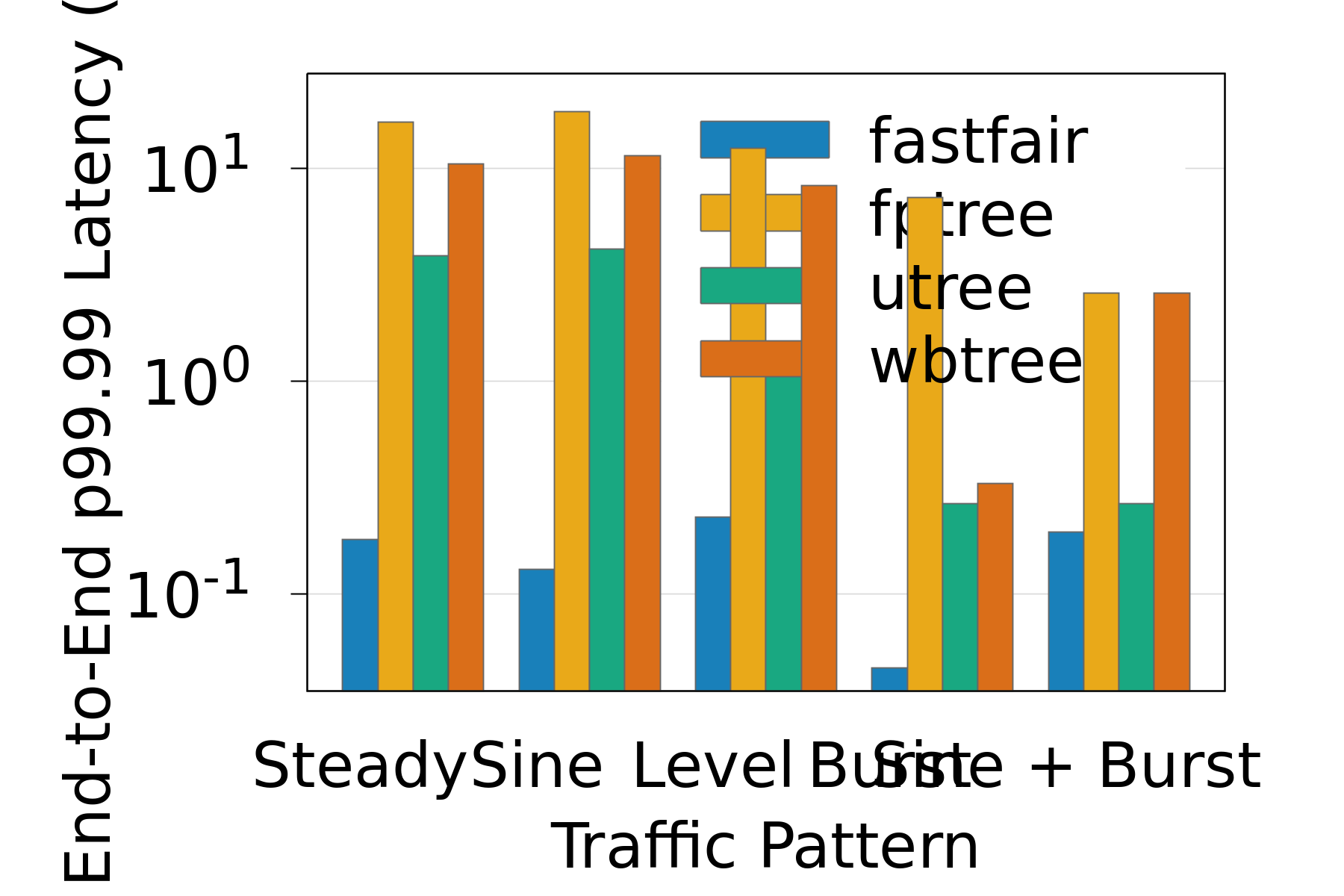}
        \caption{Demand shape changes p99.99 end-to-end latency}
        \label{fig7}
    \end{minipage}
\end{figure}
\fi

This difference indicates that the cost of introducing writes depends strongly
on the index design.  FAST\&FAIR's substantial throughput degradation may be
partly attributed to its failure-atomic shift operations, which often modify multiple key
and pointer entries within a tree node. Such modifications increase cache-line
write traffic and sensitivity to memory interference.  This increases the
index's susceptibility to cache pollution, leading to substantial performance
degradation.

%FAST\&FAIR's substantial throughput degradation results
%from its failure-atomic shift operations. As the write fraction increases, shift
%operations modify a large number of key and pointer entries within a tree node,
%which makes more cache lines dirty than other indexes. 

\subsection{Query Arrival Patterns and Tail Latency}
\label{sec:eval_arrival}

Conventional closed-loop benchmarks are useful for measuring peak throughput,
but each worker generates its next request only after the previous one
completes. This couples request generation to processing speed and prevents the
benchmark from independently controlling the offered load. AuxBench's open-loop
execution separates request generation from index processing, allowing different
indexes to be evaluated under the same externally specified arrival pattern.

\subsubsection{Queueing near Saturation}

Figure~\ref{fig6} shows query response time including queueing   delay as the
offered request rate increases. At relatively low request rates, requests can be
processed without substantial queueing.  As the offered rate approaches the
processing capacity of an index, queueing delay increases sharply while service
time remains comparatively stable. The open-loop workload exposes a latency
component that cannot be observed in a conventional saturated closed-loop
microbenchmark. Specifically, the indexes reach saturation at different offered
rates. Consequently, relatively small differences in sustainable processing
capacity can produce much larger differences in response latency near
saturation.
Thus, indexes with comparable operation latencies under isolated execution may
provide substantially different response times when subjected to the same
external request rate.

\subsubsection{Time-Varying Request Arrivals}

We further examine how the temporal distribution of requests affects
high-percentile latency. Figure~\ref{fig7} compares steady, sinusoidal,
level-shifting, bursty, and combined arrival patterns with comparable average
request rates. Despite similar average request rates, the workloads produce
substantially different p99.99 end-to-end latencies. Bursty and time-varying
arrivals can temporarily raise the offered rate above an index's processing
capacity, causing requests to accumulate in the queue.

Because different indexes have different processing capacities, they respond
differently to the same arrival pattern. An arrival rate that remains
sustainable for one index may temporarily exceed the capacity of another,
resulting in much larger tail latency. This experiment confirms that average
throughput alone is insufficient for comparing indexes under application-level
request streams.

Together, the open-loop experiments provide a complementary perspective to
throughput measurements. They reveal how differences in processing capacity
translate into response-time differences under externally controlled demand.

\subsection{Summary of Findings}
\label{sec:eval_summary}

AuxBench reveals the following important findings that were not identified in
previous studies using conventional index benchmarks.

First, {\bf cache-conscious designs do not necessarily remain competitive under
application-induced cache pressure}. Although $\mu$Tree was designed to improve
cache locality, it performs the worst among the evaluated indexes
under severe cache pollution. In contrast, FPTree maintains relatively stable
performance, suggesting that cache locality and resilience to cache interference
are distinct properties.

Second, {\bf concurrency optimizations do not always translate into
application-level performance gains}. Although indexes with fine-grained
synchronization scale well under saturated index-only workloads, their
advantages diminish considerably when application work is interleaved with index
operations. Furthermore, FAST\&FAIR, which employs failure-atomic shift
operations for persistent updates, exhibits much greater throughput degradation
under write-intensive workloads than $\mu$Tree.

Third, {\bf peak throughput does not fully characterize index performance under
externally generated requests}. Small differences in processing capacity can
lead to substantial differences in queueing delay and tail latency near
saturation, while time-varying arrivals further amplify these differences.

These findings do not contradict the results of prior studies but reveal
performance trade-offs that isolated benchmarks may overlook. By incorporating
application memory activity and request dynamics, AuxBench provides a broader
perspective on the performance of byte-addressable indexes.

\section{Conclusion and Future Work}

Conventional index benchmarks evaluate indexes in isolation, although indexes
typically operate as components of larger applications. We presented AuxBench, a
microbenchmark that addresses this limitation by independently controlling
post-lookup application work, request arrival patterns, operation mixes, and key
distributions. Our evaluation demonstrates that these workload characteristics
can substantially affect the relative performance of existing indexes, sometimes
changing their rankings and diminishing the benefits of concurrency-oriented
optimizations. These findings highlight the importance of evaluating indexes
under application-level workloads rather than relying solely on isolated,
peak-throughput measurements.

Our study focuses on how application-level memory activity affects index
performance and rankings. Intel Optane DCPMM provides a useful platform for
this investigation, although the findings may not fully generalize to
CXL-attached memory. Future work will evaluate CXL systems, validate synthetic
workloads against real-world applications, separately measure index execution
and post-lookup processing times, and use hardware performance counters to
better explain index-specific performance differences.

%\section*{Acknowledgments}
%Generative AI tools were used to assist with debugging experimental code and correcting English grammar.

\bibliographystyle{ACM-Reference-Format}
\bibliography{references}

\end{document}